\documentclass[twocolumn,prl,superscriptaddress,floatfix]{revtex4}
\usepackage{graphicx}
\usepackage{setspace}

\newcommand {\be} {\begin{eqnarray}}
\newcommand {\ee} {\end{eqnarray}}

\begin{document}

\title{Low-Energy Quasiparticles Probed by Heat Transport in the Iron Based Superconductor LaFePO}
\author{Michael Sutherland}
\affiliation{Cavendish Laboratory, University of Cambridge, J.J. Thomson Ave, Cambridge, CB3 0HE UK}
\author{J. Dunn}
\affiliation{GWPI and Department of Physics and Astronomy, University of Waterloo, Waterloo, Ontario, Canada, N2L 3G1}
\author{William Toews}
\affiliation{GWPI and Department of Physics and Astronomy, University of Waterloo, Waterloo, Ontario, Canada, N2L 3G1}
\author{Eoin O'Farrell}
\affiliation{Cavendish Laboratory, University of Cambridge, J.J. Thomson Ave, Cambridge, CB3 0HE UK}
\author{James Analytis}
\affiliation{Geballe Laboratory
for Advanced Materials and Department of Applied Physics, Stanford
University, Stanford, California 94305, USA}
\affiliation{Stanford Institute of
Energy and Materials Science, SLAC National Accelerator Laboratory, 2575
Sand Hill Road, Menlo Park
94025, California 94305, USA}
\author{Ian Fisher}
\affiliation{Geballe Laboratory
for Advanced Materials and Department of Applied Physics, Stanford
University, Stanford, California 94305, USA}
\affiliation{Stanford Institute of
Energy and Materials Science, SLAC National Accelerator Laboratory, 2575
Sand Hill Road, Menlo Park
94025,California 94305, USA}
\author{R.W. Hill}
\affiliation{GWPI and Department of Physics and Astronomy, University of Waterloo, Waterloo, Ontario, Canada, N2L 3G1}

\begin{abstract}
We have measured the thermal conductivity of the iron pnictide superconductor LaFePO down to temperatures as low as $T$=60mK and in magnetic fields up to 5~T.  The data shows a large residual contribution that is linear in temperature, consistent with the presence of low energy electronic quasiparticles.  We interpret the magnitude of the linear term, as well as the field and temperature dependence of thermal transport in several pairing scenarios. The presence of an unusual supralinear temperature dependence of the electronic thermal conductivity in zero magnetic field, and a high scattering rate with minimal $T_c$ suppression argues for a sign-changing nodal $s{\pm}$ state.

\end{abstract}

\maketitle

\section{\label{intro}Introduction}
One of the most important aspects of formulating a detailed and complete understanding of superconductivity in the iron-based superconductors is knowing the symmetry of the superconducting gap, because of its implications for the pairing mechanism.  In the presence of a weak electron-phonon interaction \cite{Boeri}, focus has centered on a pairing mechanism associated with magnetic fluctuations \cite{Mazin09}.   
The combination of a magnetic coupling mechanism and the multiband nature of the Fermi surface gives rise to the possibility of different sheets of the Fermi surface with superconducting gaps of different phase and symmetry. The permutations of these possibilities leads to a very rich phenomenology, and makes the interpretation of experimental results far more difficult than in a single band situation, as for example in the cuprates.  Moreover, theoretical calculations \cite{Graser} have shown that many of these potential gap arrangements lie very close in energy and therefore each requires careful consideration. 

Experimental studies across the various members of the pnictide family of superconductors have revealed a complex picture \cite{Johnston10}. NMR Knight shift measurements establish that the superconductivity in the iron pnictides is singlet in nature \cite{Terasaki09,Grafe08,Matano08,Yashima09}, and penetration depth measurements \cite{Hashimoto09}, thermal conductivity \cite{Tanatar10}, Andreev-reflection spectroscopy \cite{Szabo09} and Angle Resolved Photoemission (ARPES) studies \cite{Ding08}, have revealed fully gapped superconductivity in many compounds. However low temperature penetration depth measurements in BaFe$_2$(As$_{1-x}$P$_x$)$_2$ \cite{Hashimoto10} and LaFePO \cite{Fletcher09}, and Raman scattering in LaFePO \cite{Muschler09}, observe low energy excitations suggestive of a line node. Attempts to resolve these observations with theory have highlighted the possible role of intraorbital interactions \cite{Chubukov09,Thomale11} and the precise details of Fermi surface nesting \cite{Kemper10} in determining different gap topologies. It has even been shown that in the case of an isotropic sign-changing $s$-wave ($s{\pm}$) state, under certain conditions disorder can create subgap states which can give rise to a low energy density of states \cite{Golubov97} mimicking the presence of nodes. The closeness in energy of these various gap structures suggests that different materials may support different gap topologies, which ensures a rich phenomenology across the pnictide materials. Clearly, it is extremely important to conduct high-quality experiments in many compounds in order to establish whether the low-energy excitations are intrinsic (arising from nodes) or extrinsic (arising from disorder) so that scenarios for superconductivity in this multi-band system may be tested. 

To shed light on this issue, we use thermal conductivity measurements as a bulk probe of the superconducting state. By carrying out our studies at temperatures as low as 40 mK and in the presence of an applied magnetic field, we are able to accurately separate out contributions from electrons ($\kappa_{e}$) and phonons ($\kappa_{ph}$). Our study unambiguously reveals the presence of a superconducting gap with line nodes, by establishing a non-zero linear component of thermal conductivity as the temperature approaches absolute zero, which arises from low-energy quasiparticle excitations. A quantitative analysis of the magnitude of the electronic thermal conductivity and its evolution with temperature and applied magnetic field is used to attempt to discriminate between $d$-wave and nodal $s_\pm$ superconductors. 

\section{\label{expt}Experimental}

Our study focuses on LaFePO, an iron-phosphide superconductor with $T_c$ $\sim$ 7.5 K, and a reported $H_{c2}$~$\sim$~900mT \cite{Yamashita}, which can be made with very high purity. LaFePO is nonmagnetic, isostructural with LaFeAsO and has a Fermi surface consisting of almost nested electron and hole pockets \cite{Coldea08}. Our samples were small, single crystal platelets of dimensions $\sim$ 200 $\mu$m $\times$ 50 $\mu$m $\times$ 20 $\mu$m  grown via a Sn flux method \cite{Carrington09}. Electrical resistivity measurements confirmed the high quality of our samples, revealing a residual resistivity ratio of $\rho_{300K}$/$\rho_{0K}$ = 65 with $\rho_0$ = 2.4 $\mu\Omega$cm, measured by suppressing superconductivity with $H$ = 2 T applied along the $c$-axis.

Contacts with very low thermal resistance are essential for measurements of thermal conductivity at $T <$ 1K, as thermal decoupling between the electron and phonon degrees of freedom is known to potentially lead to an underestimate of $\kappa_{e}$ \cite{Smith05}. Highly conductive contacts were prepared by Ar etching the crystal surface and evaporating Pt pads, which were bonded to gold wires with silver epoxy. At low temperatures the electrical resistance of these is less than 1 m$\Omega$.

The thermal conductivity was measured using a single heater-two thermometer method. The heat current was supplied along the $ab$ planar direction direction, and the magnetic field applied perpendicular to this. The measurements were made in a dilution refrigerator by varying the temperature from 0.04~K to $> 0.7$~K at fixed magnetic field.  To ensure homogeneous flux penetration, the samples were field-cooled by cycling to $T>T_c$ before changing the field. The error in the absolute value of the conductivity is estimated to be approximately $10\%$, which is set by uncertainties in the geometric factor of the sample.  The relative error between temperature sweeps at different fields however is lower, on the order of 3\%.

\section{\label{results}RESULTS and DISCUSSION}
\begin{figure}
\begin{center}
\scalebox{0.6}{\includegraphics{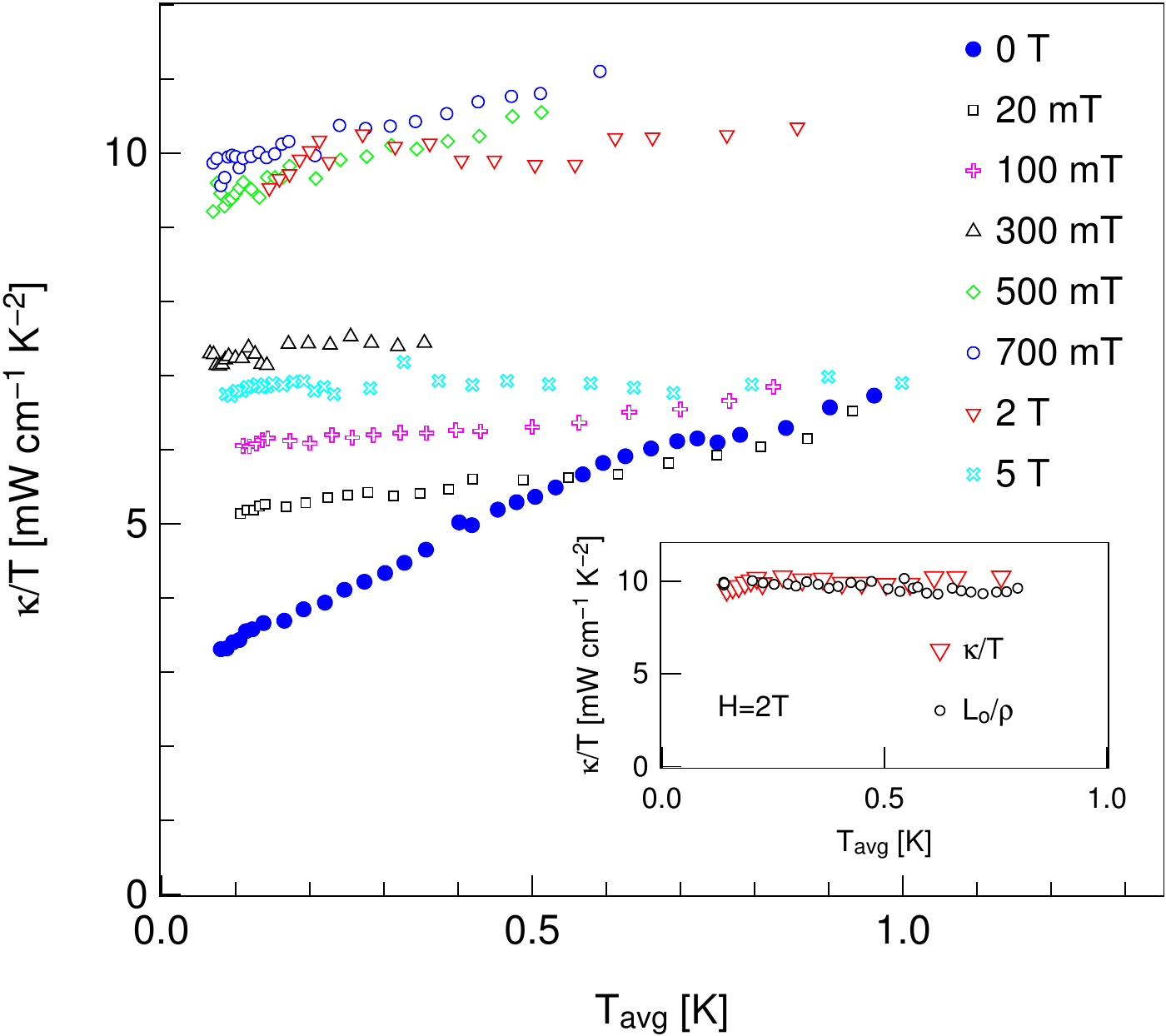}}
\caption{\label{fig:k-H}Temperature dependence of thermal conductivity divided by temperature at magnetic fields applied perpendicular to the c-axis. \emph{Inset}: Thermal conductivity divided by temperature and charge conductivity in thermal units ($L_0/\rho$) at $H$=2~T as a function of temperature.  The agreement between the two independent measurements shows that the Wiedemann Franz law is satisfied.}
\end{center}
\end{figure}


Fig. \ref{fig:k-H} shows the raw measurements of the in-plane thermal conductivity plotted as $\kappa / T$ versus average sample temperature ($T_{avg}$) for temperatures from 60 mK to 1.0 K, for a variety of magnetic fields applied parallel to the $c$-axis and perpendicular to the heat current direction. In a simple non-magnetic metal, the temperature dependence of the thermal conductivity is expected to be governed by two terms - an electronic linear term $\kappa_{e}$ $\propto T$ and a cubic phonon term  $\kappa_{ph}$ $\propto T^3$. In the superconducting state, $\kappa_{e}$ is dramatically altered as it is determined solely by thermally excited quasiparticles resulting from broken Cooper pairs. Below $T_c$,  the low-energy landscape of excitations depends sensitively on the topology of the superconducting gap.

Beginning with the zero-field data, we note that the dependence of $\kappa/T$ on temperature is much stronger than any of the in-field data, which suggests an extra temperature dependent contribution to thermal conductivity that is rapidly suppressed with field. This qualitative feature has been seen in the superconducting states of both cuprate \cite{Hill-PRL1} and filled skutterudite \cite{Hill-POS-PRL} superconductors. Since the phonon term is not expected to be particularly sensitive to field, we interpret this extra zero-field conductivity as arising from thermally excited quasiparticles, in addition to the expected linear term.   

We now set out to separate the phonon and electron contributions to $\kappa$. When a field of 20~mT or greater is applied, the temperature dependence of $\kappa/T$ is greatly reduced, and remains unchanged for higher fields. We thus assume that for $H \geq 20$ mT, phonons are the only other contribution to the conductivity in addition to the linear electronic term. The phonon term is expected to be small at low temperatures, as demonstrated by the relatively small slopes of the data for H $\geq$ 20mT in Fig. \ref{fig:k-H}. We then take as our phonon contribution the total conductivity measured at $H$ = 20 mT less the residual linear term $\kappa_0/T$ at that field (we define $\kappa_0/T$ as the limit of $\kappa/T$ as $T \rightarrow$ 0).

\be
\label{eq:phonons}
\frac{\kappa_{ph}}{T} = \frac{\kappa(20~\text{mT})}{T} - \frac{\kappa_0(20~\text{mT})}{T} 
\ee

\begin{figure}
\begin{center}
\scalebox{0.6}{\includegraphics{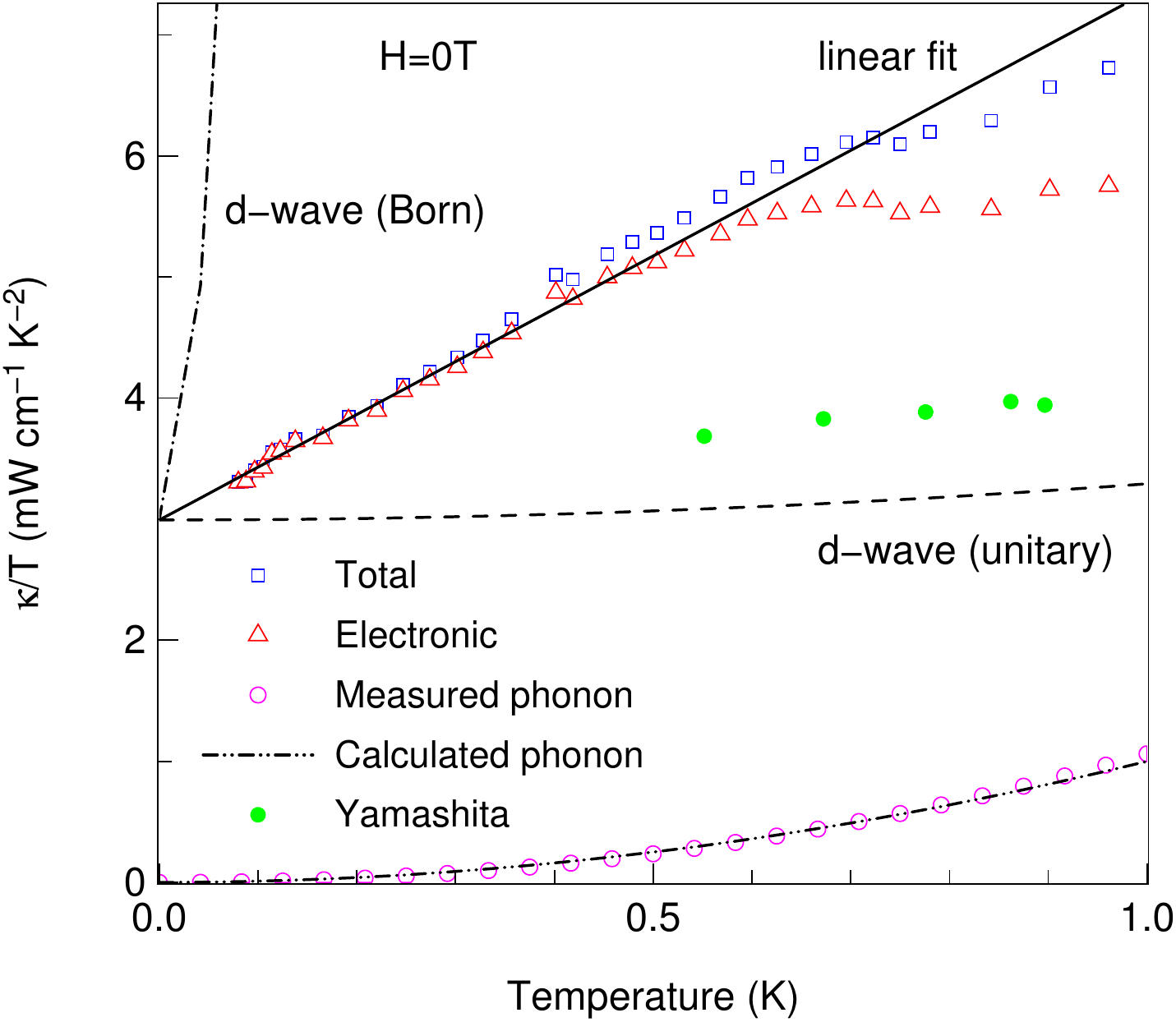}}
\caption{\label{fig:ke-kph}Thermal conductivity divided by temperature at zero magnetic field. We show the total thermal conductivity, as well as the estimated electronic and phonon contribution and the calculated electronic contribution in the d-wave Born and unitary scattering limits (described in the text). Closed circles are data from \cite{Yamashita}.}
\end{center}
\end{figure}

This phonon conductivity is plotted in Fig. \ref{fig:ke-kph}.  We assess the magnitude of $\kappa_{ph}$ by fitting to a cubic temperature dependence consistent with scattering of phonons at the boundary of the sample, giving $\kappa_{ph}$ = 1.2 $T^3$ mW/K$^4$cm.  Theoretically, the coefficient is given by kinetic theory to be equal to the product of the phonon heat capacity, $\beta$, the sound velocity, $v_s$ and a mean-free path equivalent to the geometric average of size of the sample cross-section $(a~\text{x}~b)$, $l_{ph} = \sqrt{4ab}/\pi$.

\be
\label{eq:ph-theory}
\kappa_{ph} = \frac{1}{3} \beta T^3 v_s l_{ph}
\ee

Measurements of specific heat \cite{Kohama} give a phonon contribution of $\beta$ = 0.16 mJ/K$^4$mol and we can estimate an average sound velocity of $v_s$ = 6263 m/s from the Debye temperature, $\Theta_D$ = 371 K. Combining this with our sample dimensions $l_{ph}$ = 2.7 x 10$^{-5}$ m, gives a phonon conductivity of $\kappa_{ph} = 1~ T^3$ mW/K$^4$cm which is consistent with our measured value.

We can now use this phonon conductivity to extract the entire electronic conductivity in zero field by taking
\be
\label{eq:k_e}
\frac{\kappa_{e}(H=0)}{T} = \frac{\kappa(H=0)}{T} -\frac{\kappa_{ph}}{T}
\ee

This yields the $\kappa_e/T$ curve labeled `Electronic' shown in Fig. \ref{fig:ke-kph}, which evolves linearly with temperature between 60 mK and 600 mK, making the total electronic conductivity the sum of a linear and quadratic term. When the the data is extrapolated to zero temperature, the residual linear term is $\kappa_0/T = 3.0 \pm 0.3$ mW/K$^2$cm. Extrapolating the raw total conductivity fitted in the same way over the same temperature range leads to a linear term $\kappa_0/T = 2.9 \pm 0.3$ mW/K$^2$cm, indicating the conclusion is not significantly dependent on any analysis. The magnitude of the normal state conductivity is taken as the linear term at 700 mT, $\kappa _0 (0.7$~T$)/ T$ = 9.6 $\pm$ 0.3 mW/K$^2$cm.  Consequently, the magnitude of the zero-field residual linear term is a substantial (0.31) fraction of the normal state conductivity. We note that our current study extends thermal conductivity data to almost an order of magnitude lower in temperature than previous work \cite{Yamashita}, indicated by the filled circles in Fig. \ref{fig:ke-kph}. Although there has been no attempt to subtract any phonon conductivity, the data is qualitatively and quantitatively consistent with that reported here. This allows us to make firmer conclusions about the magnitude of $\kappa_0/T$, to capture any temperature dependence of $\kappa_e/T$, and to rule out the opening of a low energy gap between 60 and 600 mK. 

Having extracted $\kappa_{e}/T(H=0)$, we may now compare our data to quantitative predictions for the size and temperature dependence of the electronic conductivity expected to arise for various gap symmetries \cite{Mishra, Bang}.

{\it Fully gapped $s_\pm$ symmetry}: Conventional thinking regarding a fully-gapped superconducting state leads to the conclusion that the residual electronic conductivity is zero because the quasiparticle excitations are exponentially suppressed in the low temperature limit. Interestingly, in the sign changing $s_\pm$ state, where isotropic gaps of different signs exist on different Fermi surface sheets, interband scattering can lead to a small but finite density of states even at zero energy \cite{Mishra}. The magnitude of the thermal conductivity associated with this band remains below 1 \% of the normal state conductivity though, even for sizable scattering rates. Consequently the data reported here for LaFePO is quantitatively inconsistent with a fully-gapped $s_\pm$ symmetry. 

{\it Nodal $s_\pm$ symmetry}:  Depending on the exact details of the magnetic fluctuations that provide the electronic coupling, the $s_\pm$ symmetry can develop minima on one of the Fermi surfaces. These minima can either be very deep, retaining the same sign of the order parameter, or may actually change signs. Many transport models treat the order parameter in an effective two band model, with an isotropic gap on the sheet centered at the $\Gamma$ point, with the sheet about the $M$ point playing host to the anisotropic Fermi surface \cite{Graser}. 

As firmly established in a $d$-wave superconductor, a superconducting gap with nodes in the presence of impurity pair-breaking leads to an impurity band of quasiparticle excitations that give rise to a residual linear term in the electronic thermal conductivity.  The essential difference between the conductivity in the nodal $s_\pm$ case and the $d$-wave case is that, in the latter, the conductivity is universal in the sense that the magnitude does not depend on the scattering rate.  This is not the case for the $s_\pm$ symmetry as shown in calculations by Mishra et al. \cite{Mishra} where the magnitude of the conductivity is strongly dependent on the normal state scattering rate.

The normal state scattering rate for the sample reported here can be estimated from the normal state electronic conductivity, $\kappa_e(N)$, using kinetic theory which gives $\kappa_e(N)$=$\gamma v_F^2\tau/3$. The average magnitude of the electronic specific heat $\gamma$ measured by two groups is $C_e /T= \gamma \sim 11.5$ mJ/K$^2$mol \cite{Kohama,McQueen08} and the Fermi velocity is $v_F = 1.5 \times 10^5$ m/s from ARPES measurements \cite{DHLu}.  This leads to a normal state scattering rate $\Gamma = 1/2\tau = 1.1 \times 10^{12}$ s$^{-1}$.  In reduced units with respect to $T_c$, this is $\hbar \Gamma / k_B T_c = 1.1$.  

An alternate method is to estimate $\Gamma$ from de Haas-van Alphen measurements of the mean free path $\ell_0$. Using $\ell_0$ $\approx$ 1000 $\AA $ \cite{Carrington09} and $v_F$ $\approx$ 1.5 $\times$ 10$^5$ m/s  \cite{DHLu} yields $\tau$ = $v_F$/$\ell_0$ = 1 $\times$ 10$^{-13}$ s. This would suggest our samples of LaFePO are in the regime $\hbar\Gamma/k_BT_c$ $\sim$ 0.8, consistent with the previous method. Similar values have been observed in other pnictide compounds, for instance $\hbar\Gamma/k_BT_c$ $\sim$ 1 - 2 has been inferred from transport and penetration depth measurements in Ba(Fe$_{1-x}$Co$_x$)$_2$As$_2$ \cite{Tanatar10}, LiFeAs \cite{Tanatar11} and Ba$_{1-x}$K$_x$Fe$_2$As$_2$ \cite{Luo09}.

This high value of scattering rate might be expected to result in significant $T_c$ suppression, however this does not appear to be the case in LaFePO. Indeed $T_c$ has been shown to be relatively insensitive to disorder, increasing $\rho_0$ by a factor of 5 has no noticeable effect on $T_c$ \cite{Analytis08}.  These observations are consistent with theoretical models describing how nodes nodes of an extended-s state may be lifted by disorder, leading to a disorder induced $T_c$ suppression much more gradual than that expected from Abrikosov-Gor'kov theory \cite{Mishra09}.

In both the `deep minima' and sign changing node scenario, a scattering rate of this order of magnitude can yield a value of $\kappa_0/T$ that is a sizable fraction of the normal state value \cite{Mishra}, although other parameters such as the relative size and number of the gap maxima are required for a detailed quantitative comparison. At finite temperatures, the extra $T^2$ contribution to $\kappa_e$ is more consistent with the sign changing nodes scenario. For a scattering rate of $\hbar\Gamma/k_BT_c$ $\sim$ 0.1, assuming similar sized gap maxima on both sheets, $\kappa_e/T$ is predicted to behave approximately linearly up to $T$=0.2 $T_c$, as in our data. For a reasonable range of parameters in the `deep minima' scenario, a constant value of  $\kappa_e/T$ is expected \cite{Mishra}, which is not reflected in our data. A detailed calculation using measured parameters for LaFePO and a full account of the details of inter versus intraband scattering would be useful in distinguishing between these scenarios.

{\it $d$-wave symmetry}:  The thermal conductivity of a $d$-wave superconductor has been studied in great detail both experimentally and theoretically.  Theory predicted a universal linear electronic conductivity with a second order term that is cubic in temperature and depends strongly on scattering rate and strength \cite{Graf}.
\be
\label{eq:ke-d-wave}
\frac{\kappa_e}{T}(T) = \left[\frac{\kappa_0}{T} + \frac{7\pi^2}{15}\left(\frac{a^2T}{\gamma}\right)^2\right]
\ee
where $\gamma$ is the impurity bandwidth and $a$ is a constant that depends on the scattering strength. 
  Both these attributes have been observed in high-temperature cuprate superconductors \cite{Taillefer, Hill-PRL1}.  In this case, we can estimate the value of the universal linear term for LaFePO using \cite{Graf}:
\be
\label{eq:k0-d-wave}
\frac{\kappa_0}{T} = \left(\frac{4}{\pi}\frac{\hbar \Gamma}{\Delta_0}\frac{1}{\mu}\right)\frac{\kappa_n}{T}
\ee
Taking the weak coupling $d$-wave gap expression, $\Delta_0 = 2.14 k_B T_c$, and that $\mu$ = 2 represents the slope of the gap at the node for a pure $d$-wave function, we compute the universal linear conductivity, $\kappa_0/T = 2.7$ mW/K$^2$cm.  This is in excellent agreement with the measured value shown in Fig. \ref{fig:ke-kph}.  Turning to the second term in Eq. \ref{eq:ke-d-wave}, the main observation is that the temperature dependence is linear, i.e. the low temperature correction to the linear term is quadratic not cubic.  To better evaluate the level of consistency, we can estimate the magnitude of the conductivity in this model using the normal state scattering rate in both the limit of strong (unitary) and weak (Born) scattering. In the unitary limit, the impurity bandwidth is given by $\gamma = 0.63 \sqrt{\Delta_0 \Gamma} = 8.5$ K and $a = 1/2$ \cite{Graf}.  This gives the curve shown in Fig. \ref{fig:ke-kph} for unitary scattering.  On the same plot, we show the expectation for the weak scattering limit computed for $a = (\pi v_2 \tau_0)/2$ and $\gamma = 4\Delta_0\exp(-\pi\Delta_0/2\Gamma)$ = 0.1 K.  

Clearly the magnitude of the measured conductivity  is bounded between these two limits indicating the possibility that the scattering phase shift lies between 0 and $\pi/2$, and consequently the range of applicability of the low temperature correction as defined by $\gamma$ will lie between 0.1 - 8.5 K.  The possibility of anisotropic scattering should also be considered as a means to reconcile experiment and theory in this case. Consequently, a comparison with the $d$-wave symmetry model provides quantitative agreement for the linear term on one hand, but a qualitative (not cubic) and quantitative discrepancy with the temperature dependence of the electronic conductivity above the linear term.

An important test of the various gap scenarios discussed above is investigating the sensitivity of $\kappa_e/T$ to scattering rate, which is expected to be negligible in the $d$-wave universal limit and be substantial in the $s_\pm$ scenarios. In the previously reported LaFePO study, we note that the normal state conductivity is 6.0 mW/K$^2$cm \cite{Yamashita}, compared to 9.6 mW/K$^2$cm in the present study. This suggests that the scattering rate for the previous study is 1.6 times higher than for this work, however the extrapolated linear term in both cases are $\sim$ 3.0 mW/K$^2$cm. Within a $d$-wave scenario, the prospect of a universal conductivity is not ruled out assuming a linear extrapolation and an error in the absolute magnitude of each measurement of ~10\% due to measurements of the geometric factor. A controlled study looking at different levels of disorder in this material would prove very useful in confirming universality and thus distinguishing between the $d$ and $s_\pm$ scenarios.


\begin{figure}
\begin{center}
\scalebox{0.6}{\includegraphics{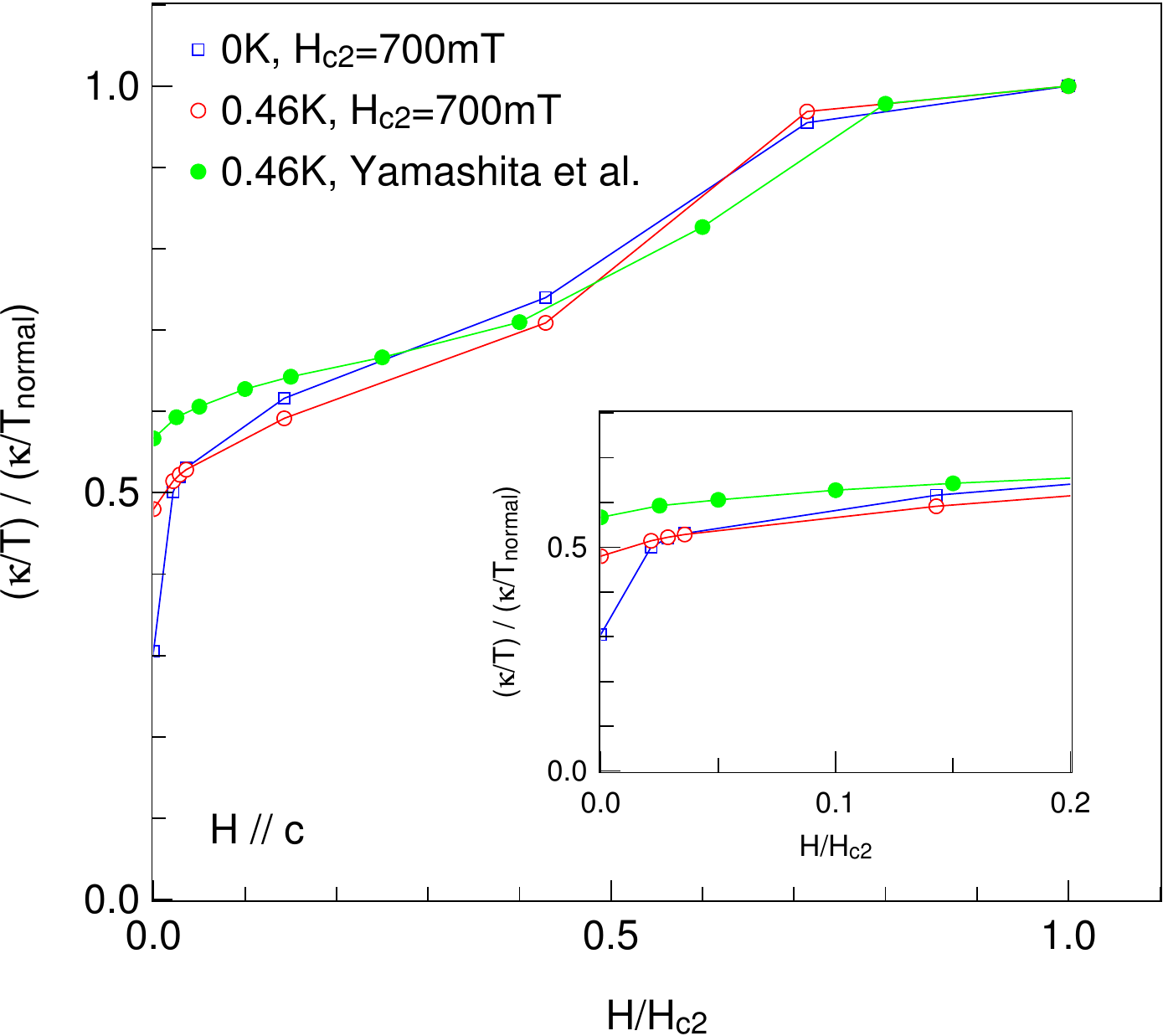}}
\caption{\label{fig:k0-H}Main panel (colour online): The linear, electronic contribution to thermal conductivity normalized to the value in the field-induced normal state. The data is plotted as a function of magnetic field in units of $H_{c2}$, with the blue squares showing the extrapolated T=0 value, and the open circles showing the T=0.46 K value. The closed circles are taken from ref \cite{Yamashita}. The inset shows a zoom of the low field region.}
\end{center}
\end{figure}


The application of a magnetic field can be used to tune a superconductor between a fully superconducting and a fully normal state, which can provide valuable information about the gap topology via the contribution of field-induced quasiparticles to thermal transport. The inset to Fig. \ref{fig:k-H} shows the normal state thermal conductivity divided by temperature, accessed by applying $H= 2$T $// c$. Also plotted is the electrical transport converted to thermal units using the Wiedemann Franz law ($L_0/\rho$).  The two data sets lie on top of one another indicating that the WF law is satisfied as it should be in a regular metal. 

At intermediate fields, the thermal conductivity rises up to meet the normal state in a striking fashion. Fig. \ref{fig:k0-H} shows the $T\rightarrow0$ extrapolated values as a function of magnetic field, where an applied field of only 20 mT \textit{doubles} $\kappa_0/T$.  We note that $\kappa/T$ continues to evolve with field up to 700mT, and remains unchanged above. This allows a convenient definition of $H_{c2}$ = 700 mT from a bulk transport probe. This is slightly below the value of $H_{c2}$ = 900 mT reported in the previous study \cite{Yamashita}.

In Fig. \ref{fig:k0-H} we also show the temperature dependence of the conductivity at $T = 0.46$ K to compare to Yamashita data.  The three curves are all quantitatively consistent having an initial steep increase, followed by a plateauing of the conductivity value before increasing again towards the normal state value.  The features are most apparent in the extrapolated zero-temperature data where the contribution is entirely electronic.  The finite temperature data are contaminated with phonons and it is clear from Fig. \ref{fig:k-H} that were the field dependence considered at a slightly higher temperature ($T> 0.5$ K) , the curve would be qualitatively different since the initial change would be a decrease in the conductivity.

In a multi-band theory, there is little difference between thermal conductivity of a $d$-wave or nodal $s_\pm$ symmetry superconductor.  In the low field regime, the nodal quasiparticles couple to the superfluid flow around a vortex and the energy states are doppler-shifted giving rise to a $\sim \sqrt{H}$ behaviour \cite{Hirshfeld}.  As the magnetic field is increased, the fully-gapped band has its gap suppressed and produces an additional channel for conductivity that causes the secondary upturn as $H_{c2}$ is approached.  The low field doppler-shift behaviour has been  observed in cuprate superconductors \cite{Chiao, Hill-PRL1}, but since these are essentially single band materials, the higher-field physics is not relevant.

In order to make a qualitative comparison with the theoretical magnetic field dependence of the conductivity reported in \cite{Mishra}, we plot the normalised zero-temperature extrapolated values of the thermal conductivity as a function of normalised field in Fig. \ref{fig:k0-H}.  There is certainly a high degree of similarity, although quantitatively, the ratio of the magnitude of the increase in conductivity effect in each of the different regimes does not appear to be consistent.  Again, tuning the various parameters may allow for a better, possibly quantitative, agreement to be found.  Beyond this, the magnetic field data supports both $d$-wave and nodal $s_\pm$ symmetries.

\section{\label{Conclusion}Conclusion}

In conclusion we have studied the temperature and field dependence of thermal transport in samples of LaFePO to temperatures as low as 60 mK. We find unambiguous evidence for low energy excitations, interpreted to arise from nodal quasiparticle excitations. A quantitative comparison to theoretical models of heat transport in the pnictide superconductors suggest our data is consistent with an $s_\pm$ nodal scenario, which better captures the linear in $T$ dependence of the zero field electronic thermal conductivity $\kappa_e/T$, and the insensitivity of $T_c$ to the measured large normal state scattering rate. Although we cannot strictly rule out a $d$-wave symmetry, we find no quantitative agreement with scattering models in either the Born or unitary limit. We suggest that a study of universality in LaFePO could help further distinguish between the two scenarios.

\section{\label{Acknowledgements}Acknowledgements}

This research was supported by NSERC of Canada and the Royal Society. Work at Stanford University was supported by the Department of Energy, Office of Basic Energy Sciences under contract DE-AC02-76SF00515. The authors wish to thank Tony Carrington and John Cooper for useful discussions, and Jos Cooper for help with making the contacts to the sample.



\bibliography{LaFePO-bib}

\begin{thebibliography}{38}
\expandafter\ifx\csname natexlab\endcsname\relax\def\natexlab#1{#1}\fi
\expandafter\ifx\csname bibnamefont\endcsname\relax
  \def\bibnamefont#1{#1}\fi
\expandafter\ifx\csname bibfnamefont\endcsname\relax
  \def\bibfnamefont#1{#1}\fi
\expandafter\ifx\csname citenamefont\endcsname\relax
  \def\citenamefont#1{#1}\fi
\expandafter\ifx\csname url\endcsname\relax
  \def\url#1{\texttt{#1}}\fi
\expandafter\ifx\csname urlprefix\endcsname\relax\def\urlprefix{URL }\fi
\providecommand{\bibinfo}[2]{#2}
\providecommand{\eprint}[2][]{\url{#2}}

\bibitem[{\citenamefont{L.Boeri et~al.}(2008)\citenamefont{L.Boeri, O.V.Dolgov,
  and A.A.Golubov}}]{Boeri}
\bibinfo{author}{\bibnamefont{L.Boeri}},
  \bibinfo{author}{\bibnamefont{O.V.Dolgov}}, \bibnamefont{and}
  \bibinfo{author}{\bibnamefont{A.A.Golubov}}, \bibinfo{journal}{Phys. Rev.
  Lett.} \textbf{\bibinfo{volume}{101}}, \bibinfo{pages}{026403}
  (\bibinfo{year}{2008}).

\bibitem[{\citenamefont{Mazin and Schmalian}(2009)}]{Mazin09}
\bibinfo{author}{\bibfnamefont{I.}~\bibnamefont{Mazin}} \bibnamefont{and}
  \bibinfo{author}{\bibfnamefont{J.}~\bibnamefont{Schmalian}},
  \bibinfo{journal}{Physica C} \textbf{\bibinfo{volume}{469}},
  \bibinfo{pages}{614} (\bibinfo{year}{2009}).

\bibitem[{\citenamefont{Graser et~al.}(2009)\citenamefont{Graser, Maier,
  Hirschfeld, and Scalapino}}]{Graser}
\bibinfo{author}{\bibfnamefont{S.}~\bibnamefont{Graser}},
  \bibinfo{author}{\bibfnamefont{T.}~\bibnamefont{Maier}},
  \bibinfo{author}{\bibfnamefont{P.}~\bibnamefont{Hirschfeld}},
  \bibnamefont{and}
  \bibinfo{author}{\bibfnamefont{D.}~\bibnamefont{Scalapino}},
  \bibinfo{journal}{New Journal of Physics} \textbf{\bibinfo{volume}{11}},
  \bibinfo{pages}{025016} (\bibinfo{year}{2009}).

\bibitem[{\citenamefont{Johnston}(2010)}]{Johnston10}
\bibinfo{author}{\bibfnamefont{D.~C.} \bibnamefont{Johnston}},
  \bibinfo{journal}{Advances in Physics} \textbf{\bibinfo{volume}{59}},
  \bibinfo{pages}{803} (\bibinfo{year}{2010}).

\bibitem[{\citenamefont{Terasaki et~al.}(2009)\citenamefont{Terasaki, Mukuda,
  Yashima, Kitaoka, Miyazawa, Shirage, Kito, Eisaki, and Iyo}}]{Terasaki09}
\bibinfo{author}{\bibfnamefont{N.}~\bibnamefont{Terasaki}},
  \bibinfo{author}{\bibfnamefont{H.}~\bibnamefont{Mukuda}},
  \bibinfo{author}{\bibfnamefont{M.}~\bibnamefont{Yashima}},
  \bibinfo{author}{\bibfnamefont{Y.}~\bibnamefont{Kitaoka}},
  \bibinfo{author}{\bibfnamefont{K.}~\bibnamefont{Miyazawa}},
  \bibinfo{author}{\bibfnamefont{P.}~\bibnamefont{Shirage}},
  \bibinfo{author}{\bibfnamefont{H.}~\bibnamefont{Kito}},
  \bibinfo{author}{\bibfnamefont{H.}~\bibnamefont{Eisaki}}, \bibnamefont{and}
  \bibinfo{author}{\bibfnamefont{A.}~\bibnamefont{Iyo}}, \bibinfo{journal}{J.
  Phys. Soc. Jpn.} \textbf{\bibinfo{volume}{78}}, \bibinfo{pages}{013701}
  (\bibinfo{year}{2009}).

\bibitem[{\citenamefont{Grafe et~al.}(2008)\citenamefont{Grafe, Paar, Lang,
  Curro, Behr, Werner, Hamann-Borrero, Hess, Leps, Klingeler et~al.}}]{Grafe08}
\bibinfo{author}{\bibfnamefont{H.-J.} \bibnamefont{Grafe}},
  \bibinfo{author}{\bibfnamefont{D.}~\bibnamefont{Paar}},
  \bibinfo{author}{\bibfnamefont{G.}~\bibnamefont{Lang}},
  \bibinfo{author}{\bibfnamefont{N.~J.} \bibnamefont{Curro}},
  \bibinfo{author}{\bibfnamefont{G.}~\bibnamefont{Behr}},
  \bibinfo{author}{\bibfnamefont{J.}~\bibnamefont{Werner}},
  \bibinfo{author}{\bibfnamefont{J.}~\bibnamefont{Hamann-Borrero}},
  \bibinfo{author}{\bibfnamefont{C.}~\bibnamefont{Hess}},
  \bibinfo{author}{\bibfnamefont{N.}~\bibnamefont{Leps}},
  \bibinfo{author}{\bibfnamefont{R.}~\bibnamefont{Klingeler}},
  \bibnamefont{et~al.}, \bibinfo{journal}{Phys. Rev. Lett.}
  \textbf{\bibinfo{volume}{101}}, \bibinfo{pages}{047003}
  (\bibinfo{year}{2008}).

\bibitem[{\citenamefont{Matano et~al.}(2008)\citenamefont{Matano, Ren, Dong,
  Sun, Zhao, and qing Zheng}}]{Matano08}
\bibinfo{author}{\bibfnamefont{K.}~\bibnamefont{Matano}},
  \bibinfo{author}{\bibfnamefont{Z.~A.} \bibnamefont{Ren}},
  \bibinfo{author}{\bibfnamefont{X.~L.} \bibnamefont{Dong}},
  \bibinfo{author}{\bibfnamefont{L.~L.} \bibnamefont{Sun}},
  \bibinfo{author}{\bibfnamefont{Z.~X.} \bibnamefont{Zhao}}, \bibnamefont{and}
  \bibinfo{author}{\bibfnamefont{G.}~\bibnamefont{qing Zheng}},
  \bibinfo{journal}{Europhys. Lett.} \textbf{\bibinfo{volume}{83}},
  \bibinfo{pages}{57001} (\bibinfo{year}{2008}).

\bibitem[{\citenamefont{Yashima et~al.}(2009)\citenamefont{Yashima, Nishimura,
  Mukuda, Kitaoka, Miyazawa, Shirage, Kihou, Kito, Eisaki, and
  Iyo}}]{Yashima09}
\bibinfo{author}{\bibfnamefont{M.}~\bibnamefont{Yashima}},
  \bibinfo{author}{\bibfnamefont{H.}~\bibnamefont{Nishimura}},
  \bibinfo{author}{\bibfnamefont{H.}~\bibnamefont{Mukuda}},
  \bibinfo{author}{\bibfnamefont{Y.}~\bibnamefont{Kitaoka}},
  \bibinfo{author}{\bibfnamefont{K.}~\bibnamefont{Miyazawa}},
  \bibinfo{author}{\bibfnamefont{P.}~\bibnamefont{Shirage}},
  \bibinfo{author}{\bibfnamefont{K.}~\bibnamefont{Kihou}},
  \bibinfo{author}{\bibfnamefont{H.}~\bibnamefont{Kito}},
  \bibinfo{author}{\bibfnamefont{H.}~\bibnamefont{Eisaki}}, \bibnamefont{and}
  \bibinfo{author}{\bibfnamefont{A.}~\bibnamefont{Iyo}}, \bibinfo{journal}{J.
  Phys. Soc. Jpn.} \textbf{\bibinfo{volume}{78}}, \bibinfo{pages}{103702}
  (\bibinfo{year}{2009}).

\bibitem[{\citenamefont{Hashimoto et~al.}(2009)\citenamefont{Hashimoto,
  Shibauchi, Kasahara, Ikada, Tonegawa, Kato, Okazaki, van~der Beek,
  Konczykowski, Takeya et~al.}}]{Hashimoto09}
\bibinfo{author}{\bibfnamefont{K.}~\bibnamefont{Hashimoto}},
  \bibinfo{author}{\bibfnamefont{T.}~\bibnamefont{Shibauchi}},
  \bibinfo{author}{\bibfnamefont{S.}~\bibnamefont{Kasahara}},
  \bibinfo{author}{\bibfnamefont{K.}~\bibnamefont{Ikada}},
  \bibinfo{author}{\bibfnamefont{S.}~\bibnamefont{Tonegawa}},
  \bibinfo{author}{\bibfnamefont{T.}~\bibnamefont{Kato}},
  \bibinfo{author}{\bibfnamefont{R.}~\bibnamefont{Okazaki}},
  \bibinfo{author}{\bibfnamefont{C.~J.} \bibnamefont{van~der Beek}},
  \bibinfo{author}{\bibfnamefont{M.}~\bibnamefont{Konczykowski}},
  \bibinfo{author}{\bibfnamefont{H.}~\bibnamefont{Takeya}},
  \bibnamefont{et~al.}, \bibinfo{journal}{Phys. Rev. Lett.}
  \textbf{\bibinfo{volume}{102}}, \bibinfo{pages}{207001}
  (\bibinfo{year}{2009}).

\bibitem[{\citenamefont{Tanatar et~al.}(2010)\citenamefont{Tanatar, Reid,
  Shakeripour, Luo, Doiron-Leyraud, Ni, Bud'ko, Canfield, Prozorov, and
  Taillefer}}]{Tanatar10}
\bibinfo{author}{\bibfnamefont{M.~A.} \bibnamefont{Tanatar}},
  \bibinfo{author}{\bibfnamefont{J.-P.} \bibnamefont{Reid}},
  \bibinfo{author}{\bibfnamefont{H.}~\bibnamefont{Shakeripour}},
  \bibinfo{author}{\bibfnamefont{X.~G.} \bibnamefont{Luo}},
  \bibinfo{author}{\bibfnamefont{N.}~\bibnamefont{Doiron-Leyraud}},
  \bibinfo{author}{\bibfnamefont{N.}~\bibnamefont{Ni}},
  \bibinfo{author}{\bibfnamefont{S.~L.} \bibnamefont{Bud'ko}},
  \bibinfo{author}{\bibfnamefont{P.~C.} \bibnamefont{Canfield}},
  \bibinfo{author}{\bibfnamefont{R.}~\bibnamefont{Prozorov}}, \bibnamefont{and}
  \bibinfo{author}{\bibfnamefont{L.}~\bibnamefont{Taillefer}},
  \bibinfo{journal}{Phys. Rev. Lett.} \textbf{\bibinfo{volume}{104}},
  \bibinfo{pages}{067002} (\bibinfo{year}{2010}).

\bibitem[{\citenamefont{Szab\'{o} et~al.}(2009)\citenamefont{Szab\'{o},
  Pribulov\'{a}, Prist\'{a}\u{s}, Bud'ko, Canfield, and Samuely}}]{Szabo09}
\bibinfo{author}{\bibfnamefont{P.}~\bibnamefont{Szab\'{o}}},
  \bibinfo{author}{\bibfnamefont{Z.}~\bibnamefont{Pribulov\'{a}}},
  \bibinfo{author}{\bibfnamefont{G.}~\bibnamefont{Prist\'{a}\u{s}}},
  \bibinfo{author}{\bibfnamefont{S.~L.} \bibnamefont{Bud'ko}},
  \bibinfo{author}{\bibfnamefont{P.~C.} \bibnamefont{Canfield}},
  \bibnamefont{and} \bibinfo{author}{\bibfnamefont{P.}~\bibnamefont{Samuely}},
  \bibinfo{journal}{Phys. Rev. B} \textbf{\bibinfo{volume}{79}},
  \bibinfo{pages}{012503} (\bibinfo{year}{2009}).

\bibitem[{\citenamefont{Ding et~al.}(2008)\citenamefont{Ding, Richard,
  Nakayama, Sugawara, Arakane, Sekiba, Takayama, Souma, Sato, Takahashi
  et~al.}}]{Ding08}
\bibinfo{author}{\bibfnamefont{H.}~\bibnamefont{Ding}},
  \bibinfo{author}{\bibfnamefont{P.}~\bibnamefont{Richard}},
  \bibinfo{author}{\bibfnamefont{K.}~\bibnamefont{Nakayama}},
  \bibinfo{author}{\bibfnamefont{K.}~\bibnamefont{Sugawara}},
  \bibinfo{author}{\bibfnamefont{T.}~\bibnamefont{Arakane}},
  \bibinfo{author}{\bibfnamefont{Y.}~\bibnamefont{Sekiba}},
  \bibinfo{author}{\bibfnamefont{A.}~\bibnamefont{Takayama}},
  \bibinfo{author}{\bibfnamefont{S.}~\bibnamefont{Souma}},
  \bibinfo{author}{\bibfnamefont{T.}~\bibnamefont{Sato}},
  \bibinfo{author}{\bibfnamefont{T.}~\bibnamefont{Takahashi}},
  \bibnamefont{et~al.}, \bibinfo{journal}{Europhys. Lett.}
  \textbf{\bibinfo{volume}{83}}, \bibinfo{pages}{47001} (\bibinfo{year}{2008}).

\bibitem[{\citenamefont{Hashimoto et~al.}(2010)\citenamefont{Hashimoto,
  Yamashita, Kasahara, Senshu, Nakata, Tonegawa, Ikada, Serafin, Carrington,
  Terashima et~al.}}]{Hashimoto10}
\bibinfo{author}{\bibfnamefont{K.}~\bibnamefont{Hashimoto}},
  \bibinfo{author}{\bibfnamefont{M.}~\bibnamefont{Yamashita}},
  \bibinfo{author}{\bibfnamefont{S.}~\bibnamefont{Kasahara}},
  \bibinfo{author}{\bibfnamefont{Y.}~\bibnamefont{Senshu}},
  \bibinfo{author}{\bibfnamefont{N.}~\bibnamefont{Nakata}},
  \bibinfo{author}{\bibfnamefont{S.}~\bibnamefont{Tonegawa}},
  \bibinfo{author}{\bibfnamefont{K.}~\bibnamefont{Ikada}},
  \bibinfo{author}{\bibfnamefont{A.}~\bibnamefont{Serafin}},
  \bibinfo{author}{\bibfnamefont{A.}~\bibnamefont{Carrington}},
  \bibinfo{author}{\bibfnamefont{T.}~\bibnamefont{Terashima}},
  \bibnamefont{et~al.}, \bibinfo{journal}{Phys. Rev. B}
  \textbf{\bibinfo{volume}{81}}, \bibinfo{pages}{220501(R)}
  (\bibinfo{year}{2010}).

\bibitem[{\citenamefont{Fletcher et~al.}(2009)\citenamefont{Fletcher, Serafin,
  Malone, Analytis, Chu, Erickson, Fisher, and Carrington}}]{Fletcher09}
\bibinfo{author}{\bibfnamefont{J.~D.} \bibnamefont{Fletcher}},
  \bibinfo{author}{\bibfnamefont{A.}~\bibnamefont{Serafin}},
  \bibinfo{author}{\bibfnamefont{L.}~\bibnamefont{Malone}},
  \bibinfo{author}{\bibfnamefont{J.~G.} \bibnamefont{Analytis}},
  \bibinfo{author}{\bibfnamefont{J.-H.} \bibnamefont{Chu}},
  \bibinfo{author}{\bibfnamefont{A.~S.} \bibnamefont{Erickson}},
  \bibinfo{author}{\bibfnamefont{I.~R.} \bibnamefont{Fisher}},
  \bibnamefont{and}
  \bibinfo{author}{\bibfnamefont{A.}~\bibnamefont{Carrington}},
  \bibinfo{journal}{Phys. Rev. Lett.} \textbf{\bibinfo{volume}{102}},
  \bibinfo{pages}{147001} (\bibinfo{year}{2009}).

\bibitem[{\citenamefont{Muschler et~al.}(2009)\citenamefont{Muschler, Prestel,
  Hackl, Devereaux, Analytis, Chu, and Fisher}}]{Muschler09}
\bibinfo{author}{\bibfnamefont{B.}~\bibnamefont{Muschler}},
  \bibinfo{author}{\bibfnamefont{W.}~\bibnamefont{Prestel}},
  \bibinfo{author}{\bibfnamefont{R.}~\bibnamefont{Hackl}},
  \bibinfo{author}{\bibfnamefont{T.~P.} \bibnamefont{Devereaux}},
  \bibinfo{author}{\bibfnamefont{J.~G.} \bibnamefont{Analytis}},
  \bibinfo{author}{\bibfnamefont{J.-H.} \bibnamefont{Chu}}, \bibnamefont{and}
  \bibinfo{author}{\bibfnamefont{I.~R.} \bibnamefont{Fisher}},
  \bibinfo{journal}{Phys. Rev. B} \textbf{\bibinfo{volume}{80}},
  \bibinfo{pages}{180510} (\bibinfo{year}{2009}).

\bibitem[{\citenamefont{Chubukov et~al.}(2009)\citenamefont{Chubukov, Vavilov,
  and Vorontsov}}]{Chubukov09}
\bibinfo{author}{\bibfnamefont{A.}~\bibnamefont{Chubukov}},
  \bibinfo{author}{\bibfnamefont{M.}~\bibnamefont{Vavilov}}, \bibnamefont{and}
  \bibinfo{author}{\bibfnamefont{A.}~\bibnamefont{Vorontsov}},
  \bibinfo{journal}{Phys. Rev. B} \textbf{\bibinfo{volume}{80}},
  \bibinfo{pages}{140515} (\bibinfo{year}{2009}).

\bibitem[{\citenamefont{Thomale et~al.}(2011)\citenamefont{Thomale, Platt,
  Hanke, and Bernevig}}]{Thomale11}
\bibinfo{author}{\bibfnamefont{R.}~\bibnamefont{Thomale}},
  \bibinfo{author}{\bibfnamefont{C.}~\bibnamefont{Platt}},
  \bibinfo{author}{\bibfnamefont{W.}~\bibnamefont{Hanke}}, \bibnamefont{and}
  \bibinfo{author}{\bibfnamefont{B.~A.} \bibnamefont{Bernevig}},
  \bibinfo{journal}{Phys. Rev. Lett.} \textbf{\bibinfo{volume}{106}},
  \bibinfo{pages}{187003} (\bibinfo{year}{2011}).

\bibitem[{\citenamefont{Kemper et~al.}(2010)\citenamefont{Kemper, Maier,
  Graser, Cheng, Hirschfeld, and Scalapino}}]{Kemper10}
\bibinfo{author}{\bibfnamefont{A.~F.} \bibnamefont{Kemper}},
  \bibinfo{author}{\bibfnamefont{T.~A.} \bibnamefont{Maier}},
  \bibinfo{author}{\bibfnamefont{S.}~\bibnamefont{Graser}},
  \bibinfo{author}{\bibfnamefont{H.-P.} \bibnamefont{Cheng}},
  \bibinfo{author}{\bibfnamefont{P.~J.} \bibnamefont{Hirschfeld}},
  \bibnamefont{and}
  \bibinfo{author}{\bibfnamefont{D.}~\bibnamefont{Scalapino}},
  \bibinfo{journal}{New J. Phys.} \textbf{\bibinfo{volume}{12}},
  \bibinfo{pages}{073020} (\bibinfo{year}{2010}).

\bibitem[{\citenamefont{Golubov and Mazin}(1997)}]{Golubov97}
\bibinfo{author}{\bibfnamefont{A.}~\bibnamefont{Golubov}} \bibnamefont{and}
  \bibinfo{author}{\bibfnamefont{I.}~\bibnamefont{Mazin}},
  \bibinfo{journal}{Phys. Rev. B} \textbf{\bibinfo{volume}{55}},
  \bibinfo{pages}{15146} (\bibinfo{year}{1997}).

\bibitem[{\citenamefont{M.Yamashita et~al.}(2009)\citenamefont{M.Yamashita,
  Nakata, Senshu, Tonegawa, Ikada, Hashimoto, Sugawara, Shibauchi, and
  Matsuda}}]{Yamashita}
\bibinfo{author}{\bibnamefont{M.Yamashita}},
  \bibinfo{author}{\bibfnamefont{N.}~\bibnamefont{Nakata}},
  \bibinfo{author}{\bibfnamefont{Y.}~\bibnamefont{Senshu}},
  \bibinfo{author}{\bibfnamefont{S.}~\bibnamefont{Tonegawa}},
  \bibinfo{author}{\bibfnamefont{K.}~\bibnamefont{Ikada}},
  \bibinfo{author}{\bibfnamefont{K.}~\bibnamefont{Hashimoto}},
  \bibinfo{author}{\bibfnamefont{H.}~\bibnamefont{Sugawara}},
  \bibinfo{author}{\bibfnamefont{T.}~\bibnamefont{Shibauchi}},
  \bibnamefont{and} \bibinfo{author}{\bibfnamefont{Y.}~\bibnamefont{Matsuda}},
  \bibinfo{journal}{Phys. Rev. B.} \textbf{\bibinfo{volume}{80}},
  \bibinfo{pages}{220509} (\bibinfo{year}{2009}).

\bibitem[{\citenamefont{Coldea et~al.}(2008)\citenamefont{Coldea, Fletcher,
  Carrington, Analytis, Bangura, Chu, Erickson, Fisher, Hussey, and
  McDonald}}]{Coldea08}
\bibinfo{author}{\bibfnamefont{A.~I.} \bibnamefont{Coldea}},
  \bibinfo{author}{\bibfnamefont{J.~D.} \bibnamefont{Fletcher}},
  \bibinfo{author}{\bibfnamefont{A.}~\bibnamefont{Carrington}},
  \bibinfo{author}{\bibfnamefont{J.~G.} \bibnamefont{Analytis}},
  \bibinfo{author}{\bibfnamefont{A.~F.} \bibnamefont{Bangura}},
  \bibinfo{author}{\bibfnamefont{J.-H.} \bibnamefont{Chu}},
  \bibinfo{author}{\bibfnamefont{A.~S.} \bibnamefont{Erickson}},
  \bibinfo{author}{\bibfnamefont{I.~R.} \bibnamefont{Fisher}},
  \bibinfo{author}{\bibfnamefont{N.~E.} \bibnamefont{Hussey}},
  \bibnamefont{and} \bibinfo{author}{\bibfnamefont{R.~D.}
  \bibnamefont{McDonald}}, \bibinfo{journal}{Phys. Rev. Lett.}
  \textbf{\bibinfo{volume}{101}}, \bibinfo{pages}{216402}
  (\bibinfo{year}{2008}).

\bibitem[{\citenamefont{Carrington et~al.}(2008)\citenamefont{Carrington,
  Coldea, Fletcher, Hussey, Andrew, Bangura, Analytis, Chu, Erickson, Fisher
  et~al.}}]{Carrington09}
\bibinfo{author}{\bibfnamefont{A.}~\bibnamefont{Carrington}},
  \bibinfo{author}{\bibfnamefont{A.}~\bibnamefont{Coldea}},
  \bibinfo{author}{\bibfnamefont{J.}~\bibnamefont{Fletcher}},
  \bibinfo{author}{\bibfnamefont{N.}~\bibnamefont{Hussey}},
  \bibinfo{author}{\bibfnamefont{C.}~\bibnamefont{Andrew}},
  \bibinfo{author}{\bibfnamefont{A.}~\bibnamefont{Bangura}},
  \bibinfo{author}{\bibfnamefont{J.}~\bibnamefont{Analytis}},
  \bibinfo{author}{\bibfnamefont{J.-H.} \bibnamefont{Chu}},
  \bibinfo{author}{\bibfnamefont{A.}~\bibnamefont{Erickson}},
  \bibinfo{author}{\bibfnamefont{I.}~\bibnamefont{Fisher}},
  \bibnamefont{et~al.}, \bibinfo{journal}{Physica C}
  \textbf{\bibinfo{volume}{469}}, \bibinfo{pages}{459} (\bibinfo{year}{2008}).

\bibitem[{\citenamefont{Smith et~al.}(2005)\citenamefont{Smith, Paglione,
  Walker, and Taillefer}}]{Smith05}
\bibinfo{author}{\bibfnamefont{M.}~\bibnamefont{Smith}},
  \bibinfo{author}{\bibfnamefont{J.}~\bibnamefont{Paglione}},
  \bibinfo{author}{\bibfnamefont{M.}~\bibnamefont{Walker}}, \bibnamefont{and}
  \bibinfo{author}{\bibfnamefont{L.}~\bibnamefont{Taillefer}},
  \bibinfo{journal}{Phys. Rev. B} \textbf{\bibinfo{volume}{71}},
  \bibinfo{pages}{014506} (\bibinfo{year}{2005}).

\bibitem[{\citenamefont{R.W.Hill et~al.}(2004)\citenamefont{R.W.Hill, Lupien,
  Sutherland, Boaknin, Hawthorn, Proust, Ronning, Taillefer, Liang, Bonn
  et~al.}}]{Hill-PRL1}
\bibinfo{author}{\bibnamefont{R.W.Hill}},
  \bibinfo{author}{\bibfnamefont{C.}~\bibnamefont{Lupien}},
  \bibinfo{author}{\bibfnamefont{M.}~\bibnamefont{Sutherland}},
  \bibinfo{author}{\bibfnamefont{E.}~\bibnamefont{Boaknin}},
  \bibinfo{author}{\bibfnamefont{D.~G.} \bibnamefont{Hawthorn}},
  \bibinfo{author}{\bibfnamefont{C.}~\bibnamefont{Proust}},
  \bibinfo{author}{\bibfnamefont{F.}~\bibnamefont{Ronning}},
  \bibinfo{author}{\bibfnamefont{L.}~\bibnamefont{Taillefer}},
  \bibinfo{author}{\bibfnamefont{R.}~\bibnamefont{Liang}},
  \bibinfo{author}{\bibfnamefont{D.~A.} \bibnamefont{Bonn}},
  \bibnamefont{et~al.}, \bibinfo{journal}{Phy. Rev. Lett.}
  \textbf{\bibinfo{volume}{92}}, \bibinfo{pages}{027001}
  (\bibinfo{year}{2004}).

\bibitem[{\citenamefont{Hill et~al.}(2008)\citenamefont{Hill, Li, Maple, and
  Taillefer}}]{Hill-POS-PRL}
\bibinfo{author}{\bibfnamefont{R.~W.} \bibnamefont{Hill}},
  \bibinfo{author}{\bibfnamefont{S.}~\bibnamefont{Li}},
  \bibinfo{author}{\bibfnamefont{M.~B.} \bibnamefont{Maple}}, \bibnamefont{and}
  \bibinfo{author}{\bibfnamefont{L.}~\bibnamefont{Taillefer}},
  \bibinfo{journal}{Phy. Rev. Lett.} \textbf{\bibinfo{volume}{101}},
  \bibinfo{pages}{237005} (\bibinfo{year}{2008}).

\bibitem[{\citenamefont{Kohama et~al.}(2008)\citenamefont{Kohama, Kamihara,
  Kawaji, Atake, Hirano, and Hosono}}]{Kohama}
\bibinfo{author}{\bibfnamefont{Y.}~\bibnamefont{Kohama}},
  \bibinfo{author}{\bibfnamefont{Y.}~\bibnamefont{Kamihara}},
  \bibinfo{author}{\bibfnamefont{H.}~\bibnamefont{Kawaji}},
  \bibinfo{author}{\bibfnamefont{T.}~\bibnamefont{Atake}},
  \bibinfo{author}{\bibfnamefont{M.}~\bibnamefont{Hirano}}, \bibnamefont{and}
  \bibinfo{author}{\bibfnamefont{H.}~\bibnamefont{Hosono}},
  \bibinfo{journal}{J. Phys. Soc. Jap.} \textbf{\bibinfo{volume}{77}},
  \bibinfo{pages}{094715} (\bibinfo{year}{2008}).

\bibitem[{\citenamefont{Mishra et~al.}(2009{\natexlab{a}})\citenamefont{Mishra,
  Vorontsov, Hirschfeld, and Vekhter}}]{Mishra}
\bibinfo{author}{\bibfnamefont{V.}~\bibnamefont{Mishra}},
  \bibinfo{author}{\bibfnamefont{A.}~\bibnamefont{Vorontsov}},
  \bibinfo{author}{\bibfnamefont{P.}~\bibnamefont{Hirschfeld}},
  \bibnamefont{and} \bibinfo{author}{\bibfnamefont{I.}~\bibnamefont{Vekhter}},
  \bibinfo{journal}{Phys. Rev. B.} \textbf{\bibinfo{volume}{80}},
  \bibinfo{pages}{224525} (\bibinfo{year}{2009}{\natexlab{a}}).

\bibitem[{\citenamefont{Bang}(2010)}]{Bang}
\bibinfo{author}{\bibfnamefont{Y.}~\bibnamefont{Bang}}, \bibinfo{journal}{Phys.
  Rev. Lett.} \textbf{\bibinfo{volume}{104}}, \bibinfo{pages}{217001}
  (\bibinfo{year}{2010}).

\bibitem[{\citenamefont{McQueen et~al.}(2008)\citenamefont{McQueen, Regulacio,
  Williams, Huang, Lynn, Hor, West, Green, and Cava}}]{McQueen08}
\bibinfo{author}{\bibfnamefont{T.~M.} \bibnamefont{McQueen}},
  \bibinfo{author}{\bibfnamefont{M.}~\bibnamefont{Regulacio}},
  \bibinfo{author}{\bibfnamefont{A.~J.} \bibnamefont{Williams}},
  \bibinfo{author}{\bibfnamefont{Q.}~\bibnamefont{Huang}},
  \bibinfo{author}{\bibfnamefont{J.~W.} \bibnamefont{Lynn}},
  \bibinfo{author}{\bibfnamefont{Y.~S.} \bibnamefont{Hor}},
  \bibinfo{author}{\bibfnamefont{D.~V.} \bibnamefont{West}},
  \bibinfo{author}{\bibfnamefont{M.~A.} \bibnamefont{Green}}, \bibnamefont{and}
  \bibinfo{author}{\bibfnamefont{R.~J.} \bibnamefont{Cava}},
  \bibinfo{journal}{Phys. Rev. B} \textbf{\bibinfo{volume}{78}},
  \bibinfo{pages}{024521} (\bibinfo{year}{2008}).

\bibitem[{\citenamefont{Lu et~al.}(2008)\citenamefont{Lu, Yi, Mo, Erickson,
  Analytis, Chu, Singh, Hussain, Geballe, Fisher et~al.}}]{DHLu}
\bibinfo{author}{\bibfnamefont{D.}~\bibnamefont{Lu}},
  \bibinfo{author}{\bibfnamefont{M.}~\bibnamefont{Yi}},
  \bibinfo{author}{\bibfnamefont{S.-K.} \bibnamefont{Mo}},
  \bibinfo{author}{\bibfnamefont{A.~S.} \bibnamefont{Erickson}},
  \bibinfo{author}{\bibfnamefont{J.}~\bibnamefont{Analytis}},
  \bibinfo{author}{\bibfnamefont{J.-H.} \bibnamefont{Chu}},
  \bibinfo{author}{\bibfnamefont{D.~J.} \bibnamefont{Singh}},
  \bibinfo{author}{\bibfnamefont{Z.}~\bibnamefont{Hussain}},
  \bibinfo{author}{\bibfnamefont{T.~H.} \bibnamefont{Geballe}},
  \bibinfo{author}{\bibfnamefont{I.~R.} \bibnamefont{Fisher}},
  \bibnamefont{et~al.}, \bibinfo{journal}{Nature}
  \textbf{\bibinfo{volume}{455}}, \bibinfo{pages}{81} (\bibinfo{year}{2008}).

\bibitem[{\citenamefont{Tanatar et~al.}(2011)\citenamefont{Tanatar, Reid,
  de~Cotret, Doiron-Leyraud, Lalibert\'{e}, Hassinger, Chang, Kim, Cho, Song
  et~al.}}]{Tanatar11}
\bibinfo{author}{\bibfnamefont{M.~A.} \bibnamefont{Tanatar}},
  \bibinfo{author}{\bibfnamefont{J.-P.} \bibnamefont{Reid}},
  \bibinfo{author}{\bibfnamefont{S.~R.} \bibnamefont{de~Cotret}},
  \bibinfo{author}{\bibfnamefont{N.}~\bibnamefont{Doiron-Leyraud}},
  \bibinfo{author}{\bibfnamefont{F.}~\bibnamefont{Lalibert\'{e}}},
  \bibinfo{author}{\bibfnamefont{E.}~\bibnamefont{Hassinger}},
  \bibinfo{author}{\bibfnamefont{J.}~\bibnamefont{Chang}},
  \bibinfo{author}{\bibfnamefont{H.}~\bibnamefont{Kim}},
  \bibinfo{author}{\bibfnamefont{K.}~\bibnamefont{Cho}},
  \bibinfo{author}{\bibfnamefont{Y.~J.} \bibnamefont{Song}},
  \bibnamefont{et~al.}, \bibinfo{journal}{Phys. Rev. B}
  \textbf{\bibinfo{volume}{84}}, \bibinfo{pages}{054507}
  (\bibinfo{year}{2011}).

\bibitem[{\citenamefont{Luo et~al.}(2009)\citenamefont{Luo, Tanatar, Reid,
  Shakeripour, Doiron-Leyraud, Ni, Bud'ko, Canfield, Luo, Wang et~al.}}]{Luo09}
\bibinfo{author}{\bibfnamefont{X.~G.} \bibnamefont{Luo}},
  \bibinfo{author}{\bibfnamefont{M.~A.} \bibnamefont{Tanatar}},
  \bibinfo{author}{\bibfnamefont{J.-P.} \bibnamefont{Reid}},
  \bibinfo{author}{\bibfnamefont{H.}~\bibnamefont{Shakeripour}},
  \bibinfo{author}{\bibfnamefont{N.}~\bibnamefont{Doiron-Leyraud}},
  \bibinfo{author}{\bibfnamefont{N.}~\bibnamefont{Ni}},
  \bibinfo{author}{\bibfnamefont{S.~L.} \bibnamefont{Bud'ko}},
  \bibinfo{author}{\bibfnamefont{P.~C.} \bibnamefont{Canfield}},
  \bibinfo{author}{\bibfnamefont{H.}~\bibnamefont{Luo}},
  \bibinfo{author}{\bibfnamefont{Z.}~\bibnamefont{Wang}}, \bibnamefont{et~al.},
  \bibinfo{journal}{Phys. Rev. B} \textbf{\bibinfo{volume}{80}},
  \bibinfo{pages}{140503(R)} (\bibinfo{year}{2009}).

\bibitem[{\citenamefont{Analytis et~al.}()\citenamefont{Analytis, Chu,
  Erickson, Kucharczyk, Serafin, Carrington, Cox, Kauzlarich, Hope, and
  Fisher}}]{Analytis08}
\bibinfo{author}{\bibfnamefont{J.~G.} \bibnamefont{Analytis}},
  \bibinfo{author}{\bibfnamefont{J.-H.} \bibnamefont{Chu}},
  \bibinfo{author}{\bibfnamefont{A.~S.} \bibnamefont{Erickson}},
  \bibinfo{author}{\bibfnamefont{C.}~\bibnamefont{Kucharczyk}},
  \bibinfo{author}{\bibfnamefont{A.}~\bibnamefont{Serafin}},
  \bibinfo{author}{\bibfnamefont{A.}~\bibnamefont{Carrington}},
  \bibinfo{author}{\bibfnamefont{C.}~\bibnamefont{Cox}},
  \bibinfo{author}{\bibfnamefont{S.~M.} \bibnamefont{Kauzlarich}},
  \bibinfo{author}{\bibfnamefont{H.}~\bibnamefont{Hope}}, \bibnamefont{and}
  \bibinfo{author}{\bibfnamefont{I.~R.} \bibnamefont{Fisher}},
  \eprint{arXiv:0810.5368}.

\bibitem[{\citenamefont{Mishra et~al.}(2009{\natexlab{b}})\citenamefont{Mishra,
  Boyd, Graser, Maier, Hirschfeld, and Scalapino}}]{Mishra09}
\bibinfo{author}{\bibfnamefont{V.}~\bibnamefont{Mishra}},
  \bibinfo{author}{\bibfnamefont{G.}~\bibnamefont{Boyd}},
  \bibinfo{author}{\bibfnamefont{S.}~\bibnamefont{Graser}},
  \bibinfo{author}{\bibfnamefont{T.}~\bibnamefont{Maier}},
  \bibinfo{author}{\bibfnamefont{P.~J.} \bibnamefont{Hirschfeld}},
  \bibnamefont{and} \bibinfo{author}{\bibfnamefont{D.~J.}
  \bibnamefont{Scalapino}}, \bibinfo{journal}{Phys. Rev. B}
  \textbf{\bibinfo{volume}{79}}, \bibinfo{pages}{094512}
  (\bibinfo{year}{2009}{\natexlab{b}}).

\bibitem[{\citenamefont{M.J.Graf et~al.}(1996)\citenamefont{M.J.Graf, Yip,
  Sauls, and Rainer}}]{Graf}
\bibinfo{author}{\bibnamefont{M.J.Graf}}, \bibinfo{author}{\bibfnamefont{S.-K.}
  \bibnamefont{Yip}}, \bibinfo{author}{\bibfnamefont{J.~A.}
  \bibnamefont{Sauls}}, \bibnamefont{and}
  \bibinfo{author}{\bibfnamefont{D.}~\bibnamefont{Rainer}},
  \bibinfo{journal}{Phys. Rev. B.} \textbf{\bibinfo{volume}{53}},
  \bibinfo{pages}{15147} (\bibinfo{year}{1996}).

\bibitem[{\citenamefont{Taillefer et~al.}(1997)\citenamefont{Taillefer,
  Lussier, Gagnon, Behnia, and Aubin}}]{Taillefer}
\bibinfo{author}{\bibfnamefont{L.}~\bibnamefont{Taillefer}},
  \bibinfo{author}{\bibfnamefont{B.}~\bibnamefont{Lussier}},
  \bibinfo{author}{\bibfnamefont{R.}~\bibnamefont{Gagnon}},
  \bibinfo{author}{\bibfnamefont{K.}~\bibnamefont{Behnia}}, \bibnamefont{and}
  \bibinfo{author}{\bibfnamefont{H.}~\bibnamefont{Aubin}},
  \bibinfo{journal}{Phys. Rev. Lett.} \textbf{\bibinfo{volume}{79}},
  \bibinfo{pages}{483} (\bibinfo{year}{1997}).

\bibitem[{\citenamefont{P.J.Hirschfeld
  et~al.}(1993)\citenamefont{P.J.Hirschfeld, W.O.Putikka, and
  D.J.Scalapino}}]{Hirshfeld}
\bibinfo{author}{\bibnamefont{P.J.Hirschfeld}},
  \bibinfo{author}{\bibnamefont{W.O.Putikka}}, \bibnamefont{and}
  \bibinfo{author}{\bibnamefont{D.J.Scalapino}}, \bibinfo{journal}{Phys. Rev.
  Lett.} \textbf{\bibinfo{volume}{71}}, \bibinfo{pages}{3705}
  (\bibinfo{year}{1993}).

\bibitem[{\citenamefont{Chiao et~al.}(1999)\citenamefont{Chiao, Hill, Lupien,
  Popic, Gagnon, and Taillefer}}]{Chiao}
\bibinfo{author}{\bibfnamefont{M.}~\bibnamefont{Chiao}},
  \bibinfo{author}{\bibfnamefont{R.~W.} \bibnamefont{Hill}},
  \bibinfo{author}{\bibfnamefont{C.}~\bibnamefont{Lupien}},
  \bibinfo{author}{\bibfnamefont{B.}~\bibnamefont{Popic}},
  \bibinfo{author}{\bibfnamefont{R.}~\bibnamefont{Gagnon}}, \bibnamefont{and}
  \bibinfo{author}{\bibfnamefont{L.}~\bibnamefont{Taillefer}},
  \bibinfo{journal}{Phys. Rev. Lett.} \textbf{\bibinfo{volume}{82}},
  \bibinfo{pages}{2943} (\bibinfo{year}{1999}).

\end{thebibliography}

\end{document}